\documentclass[aip,pop,reprint]{revtex4-1}

\usepackage{amsmath}
\usepackage{graphicx}

\newcommand{\dee}{\textnormal{d}}

\newcommand{\mb}[1]{\mathbf{#1}}

\begin{document}

\title{Cylindrical plasmas generated by an annular beam of ultraviolet light}
\date{\today}
\author{D.\ M.\ Thomas}
\thanks{Corresponding author}
\email{dmt107@imperial.ac.uk}
\affiliation{Blackett Laboratory, Imperial College London, Prince Consort Road, London, SW7 2BW, UK}
\author{J.\ E.\ Allen}
\email{John.Allen@maths.ox.ac.uk}
\affiliation{University College, University of Oxford, Oxford, OX1 4BH, UK; OCIAM, Mathematical Institute, University of Oxford, Oxford, OX2 6GG, UK}
\affiliation{Blackett Laboratory, Imperial College London, Prince Consort Road, London, SW7 2BW, UK}

\begin{abstract}
We investigate a cylindrical plasma system with ionization, by an annular beam of ultraviolet light, taking place only in the cylinder's outer region.
In the steady state both the outer and inner regions contain a plasma, with that in the inner region being uniform and field-free.
At the interface between the two regions there is an infinitesimal jump in ion density, the magnitude approaching zero in the quasi-neutral $\left( \lambda_D \rightarrow 0 \right)$ limit.
The system offers the possibility of producing a uniform stationary plasma in the laboratory, hitherto obtained only with thermally produced alkali plasmas.
\end{abstract}

\maketitle

\section{Introduction}

In two recent papers we have presented models of a plasma split into two regions of equal width, with ionization by ultraviolet light taking place only in the inner region and not in the outer \cite{Franklin13,Benilov14}.
At the interface between the two regions a double layer was formed, the ion velocity entering the outer plasma being greater than the Bohm (or ion-acoustic) velocity.
In the present paper we study the inverse case, i.e.\ ionization due to photo-violet irradiation only in the outer region. A uniform plasma fills the inner region, but instead of a double layer at the interface between the regions there is an infinitesimal jump in ion density; quasi-neutrality then holds up to the formation of a sheath at the wall. In this case the ion velocity on entering the wall sheath is about equal to the Bohm velocity. The model predicts a uniform stationary plasma in the inner region. Such a state of plasma is extremely rare; the only case known to us is that of a thermally produced alkali plasma \cite{Phelps76}. We can note that these two different situations have the common feature that the ions and electrons, which constitute the plasma, have entered the plasma from outside.
The system offers the possibility of interesting experimental work, since it is difficult to make measurements on thermally produced plasmas.

\section{Theoretical model}

\begin{figure}
\includegraphics[width=0.48\textwidth]{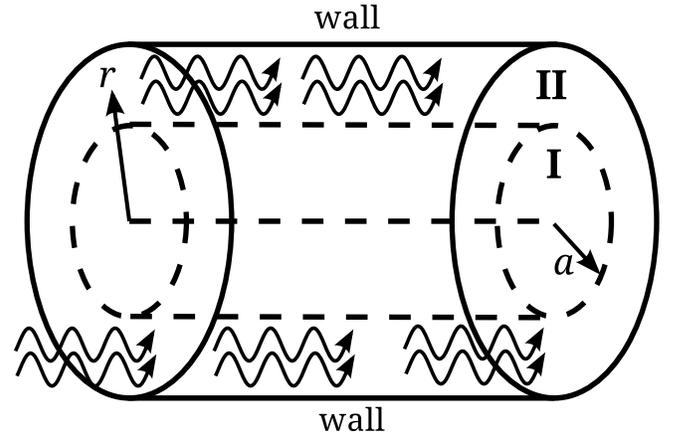}
\caption{\label{fig:sys} The cylindrical system we analyze. The system comprises two regions of plasma, the central region I ($0 \le r < a$) and the outer region II ($a \le r \le 2a$). Generation by photoionization occurs in region II with intensity $g_{II}$, and not at all in region I.}
\end{figure}

Figure \ref{fig:sys} illustrates the plasma modelled here, which four steady-state equations describe. Two are fluid equations, namely the continuity equation
\begin{equation}
\nabla \cdot (n_i \mb{v}) = g
\label{eq:unnormcont}
\end{equation}
and the ion momentum equation
\begin{equation}
m_i n_i (\mb{v} \cdot \nabla) \mb{v} + m_i \mb{v} g = - n_i e \nabla V
\end{equation}
while the other two equations are Poisson's equation
\begin{equation}
\nabla^2 V = \frac{e}{\varepsilon_0} (n_e - n_i)
\end{equation}
and the Boltzmann relation for electrons
\begin{equation}
n_e = n_0 \, \exp \left( \frac{eV}{k_B T_e} \right)
\label{eq:BR}
\end{equation}
where $n_i$ and $n_e$ are the ion and electron densities respectively, $\mb{v}$ is the ion velocity, $g$ the net generation rate, $m_i$ the ion mass, $V$ the electric potential, $T_e$ the electron temperature, and $n_0$ the electron density at the system's centre. The generation rate $g$ is nil in the inner region I but equals $g_{II} > 0$ in the outer region II.

Substituting the Boltzmann relation into the other equations reduces the system to three equations in three unknowns ($n_i$, $\mb{v}$, and $V$). Doing so, and applying the $\nabla$ operator's definition in cylindrical coordinates,
\begin{equation}
\frac{\dee (n_i v)}{\dee r} + \frac{n_i v}{r} = g
\end{equation}
\begin{equation}
m_i n_i v \frac{\dee v}{\dee r} + m_i v g = -n_i e \frac{\dee V}{\dee r}
\end{equation}
\begin{equation}
\frac{\dee^2 V}{\dee r^2} + \frac{1}{r} \frac{\dee V}{\dee r}
= \frac{e}{\varepsilon_0} \left( n_0 \exp \left( \frac{eV}{k_B T_e} \right) - n_i \right)
\end{equation}
using the fact that, by cylindrical symmetry, $\mb{v} = v \mb{\hat{r}}$ and $\nabla V = (\dee V / \dee r) \mb{\hat{r}}$.

Supplementing these equations with boundary conditions determines specific solutions.
Cylindrical symmetry dictates that all vector observables, including
$n_i \mb{v}$ and $\nabla V$, vanish at $r = 0$.
$V$ is zero too at $r = 0$, for otherwise the definition of $n_0$ would be inconsistent with eq.\ (\ref{eq:BR}). The outermost boundary condition comes from the demand that solutions be steady states: when the system is at a steady state, ions and electrons must leave through the outer wall at equal rates, so at $r = 2a$
\begin{equation}
n_i v = n_0 \exp \left( \frac{eV}{k_B T_e} \right) \sqrt{\frac{k_B T_e}{2\pi m_e}}
\end{equation}
where $2a$ is the radius of the entire system. At the interface at $r = a$ we require continuity of $V$ (a feature of models employing the electron Boltzmann relation) and the ion flux $n_i v$.

We nondimensionalize the equations with several characteristic quantities: a generation intensity, $g_0$; the Bohm speed, $c_s \equiv (k_B T_e / m_i)^{1/2}$; an ionization length, $L_i \equiv n_0 c_s / g_0$; and an electric potential, $V_T \equiv k_B T_e / e$. We may then rewrite the above equations in normalized form with the dimensionless quantities $U \equiv v / c_s$, $G \equiv g / g_0$, $R \equiv r/L_i$, $A \equiv a/L_i$, $\Phi = -V/V_T$, and $N \equiv n_i / n_0$:
\begin{equation}
\frac{\dee (NU)}{\dee R} + \frac{NU}{R} = G
\label{eq:normcont}
\end{equation}
\begin{equation}
U \frac{\dee U}{\dee R} + G \frac{U}{N} = \frac{\dee \Phi}{\dee R}
\label{eq:normmom}
\end{equation}
\begin{equation}
\frac{\dee^2 \Phi}{\dee R^2} + \frac{1}{R} \frac{\dee \Phi}{\dee R} = \frac{N - \exp(-\Phi)}{\Lambda^2}
\label{eq:normPois}
\end{equation}
where $\Lambda \equiv \lambda_D / L_i$, $\lambda_D$ being the plasma's (electron) Debye length. The boundary conditions are then
\begin{equation}
NU = \frac{\dee \Phi}{\dee R} = \Phi = 0
\label{eq:centrebc}
\end{equation}
at $R = 0$, continuity of $\Phi$ and $NU$ at $R = A$, and at the wall at $R = 2A$
\begin{equation}
NU = \sqrt{\frac{m_i}{2\pi m_e}} \exp(-\Phi)
\label{eq:equfluxbc}
\end{equation}

The continuity of $NU$ at $R = A$ does not imply continuity of $N$ or $U$ individually. A priori, $N$ and $U$ may have jump discontinuities at $R = A$. It is similarly important to note that in the quasi-neutral approximation, which takes $\Lambda \rightarrow 0$ and $n_i = n_e$, eq.\ (\ref{eq:normPois}) becomes superfluous and the wall sheath, being infinitely thin, coincides with the wall itself; the $R = 2A$ boundary condition is then given by the singularity at which $\dee U / \dee R \rightarrow +\infty$ and $U \rightarrow 1$, the (normalized) Bohm velocity.

\section{Solution for region I}

The inner region's solution is trivial because there $G = 0$, reducing the continuity and momentum equations to
\begin{equation}
\frac{\dee (NU)}{\dee R} + \frac{NU}{R} = 0
\end{equation}
and
\begin{equation}
U \frac{\dee U}{\dee R} = \frac{\dee \Phi}{\dee R}
\end{equation}
respectively.
Given eq.\ (\ref{eq:centrebc}) at $R = 0$, the continuity equation's solution is $NU = 0$ everywhere in region I, and invoking the physical condition that $N$ be everywhere finite implies $U = 0$ throughout region I.
With $U$ constantly zero, the momentum equation means $\Phi$ is constant in region I --- so $\Phi = 0$ everywhere in region I, not only at $R = 0$.
From the Boltzmann relation we then have $N = 1$ through all of region I.
Every variable is therefore constant in region I: $N = 1$ and $U = \Phi = 0$.

\section{Quasi-neutral solution for region II}

The outer region II, where generation takes place, is rather less tractable. There appears to be no closed-form solution for eqs.\ (\ref{eq:normcont})--(\ref{eq:normPois}) here. We may simplify things by assuming quasi-neutrality, where we take $n_i = n_e$ and replace eq.\ (\ref{eq:normPois}) with $N = \exp(-\Phi)$. This leaves two equations to solve for $N$ and $U$. Eq.\ (\ref{eq:normcont}), the continuity equation, is unchanged but the ion momentum equation becomes
\begin{equation}
U \frac{\dee U}{\dee R} + G \frac{U}{N} = - \frac{1}{N} \frac{\dee N}{\dee R}
\label{eq:ionmo1}
\end{equation}

As in region I the continuity equation has a simple general solution for $NU$. With $NU$ continuous at the interface of the two regions, region I's solution implies $NU = 0$ at $R = A$ and the specific solution
\begin{equation}
NU = \frac{G_{II}}{2} \frac{R^2 - A^2}{R}
\label{eq:NUII}
\end{equation}
in region II.
The natural choice of $g_0$ is $g_{II}$, making $G_{II} = 1$ and
\begin{equation}
N = \frac{1}{U} \frac{R^2 - A^2}{2 R}
\label{eq:NqnII}
\end{equation}
Substituting into eq.\ (\ref{eq:ionmo1}),
\begin{equation}
\left( \frac{1}{U} - U \right) \frac{\dee U}{\dee R}
= \frac{2 R^2 U^2 + R^2 + A^2}{R (R^2 - A^2)}
\label{eq:UqnII}
\end{equation}
solving which gives $U$ as a function of $A$ and $R$.
Again there is no closed-form solution, but we can make some inferences about the solution's behaviour. Rearranging,
\begin{equation}
\frac{\dee U}{\dee R}
= \frac{U}{1 - U^2} \frac{2 R^2 U^2 + R^2 + A^2}{R (R^2 - A^2)}
\label{eq:dUqnII}
\end{equation}
which is evidently positive where $R > A$ and $0 < U <1$.
At $R = A$ itself, $\dee U / \dee R = 1$, as shown by rearranging eq.\ (\ref{eq:NqnII}) for $U$, substituting into eq.\ (\ref{eq:dUqnII})'s right-hand side, and observing that $N = \exp(0) = 1$ at $R = A$, because $\Phi = 0$ in region I and is continuous at $R = A$.

As such $U$ strictly monotonically increases with $R$ until $U=1$, at which point $\dee U / \dee R \rightarrow +\infty$ and there is a singularity. We thus reproduce the usual quasi-neutral behaviour: the ion speed tends to the Bohm speed \cite{Bohm49} as the electric field tends to infinity.

For want of an analytic solution we solve eq.\ (\ref{eq:UqnII}) numerically. To accelerate the numerical computation we exploit the fact that given $U = U_0$ at some position, eq.\ (\ref{eq:UqnII}) may be approximated as
\begin{equation}
\left( \frac{1}{U} - U \right) \frac{\dee U}{\dee R}
= \frac{2 R^2 U_0^2 + R^2 + A^2}{R (R^2 - A^2)}
\end{equation}
where $U$ remains a function of $R$, but $U_0$ is constant. Applying the condition $U = U_0$ at $R = R_0$, this (now separable) differential equation has the solution
\begin{equation}
U = \sqrt{- W \left(
- \frac{R_0^2}{R^2} U_0^2 \exp \left( -U_0^2 \right)
\left( \frac{A^2 - R^2}{A^2 - R_0^2} \right)^{2 \left( 1 + U_0^2 \right)}
\right)}
\label{eq:apprqnU4}
\end{equation}
where $W$ is the Lambert W function's principal branch. This solution is valid for any interval $[R_0, R]$ over which $U \approx U_0$. We may then obtain the quasi-neutral solution over all of region II by chopping region II into subregions, all so narrow that $U$ is virtually constant within each, and applying eq.\ (\ref{eq:apprqnU4}) to each subregion in turn to compute $U$.

\begin{figure*}
\includegraphics[width=0.49\textwidth]{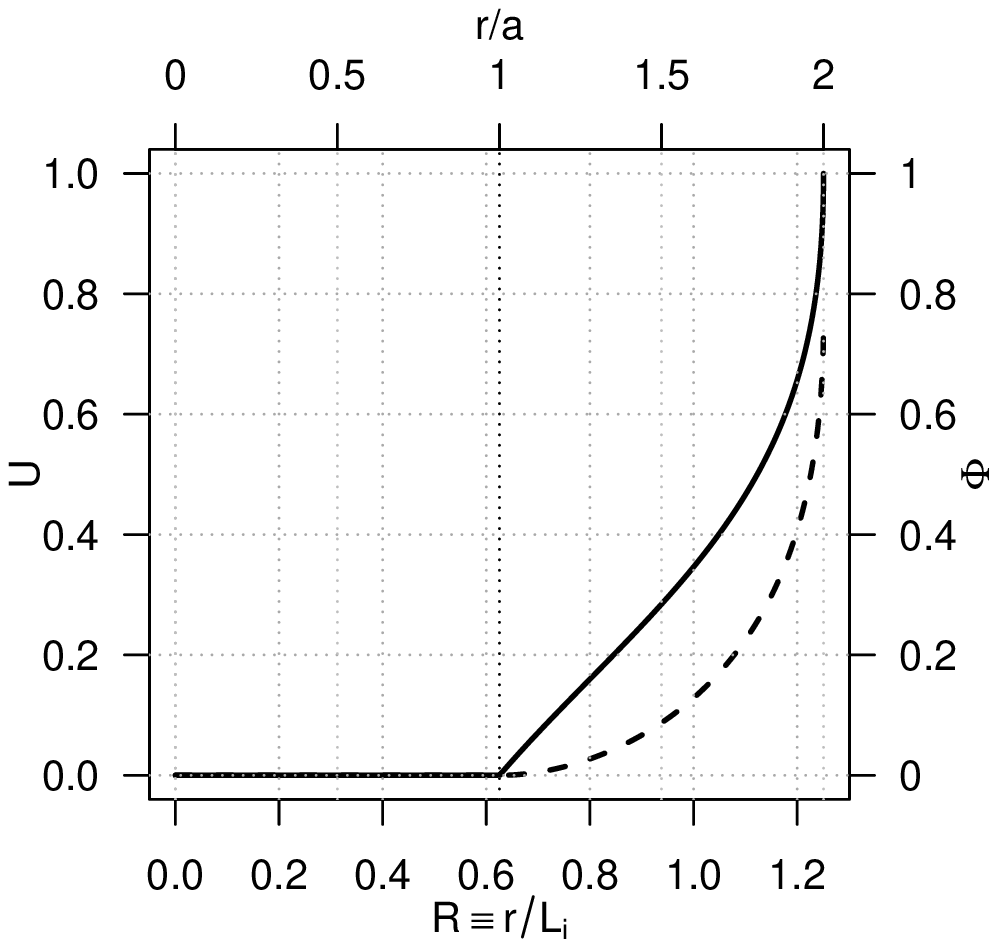}
\includegraphics[width=0.49\textwidth]{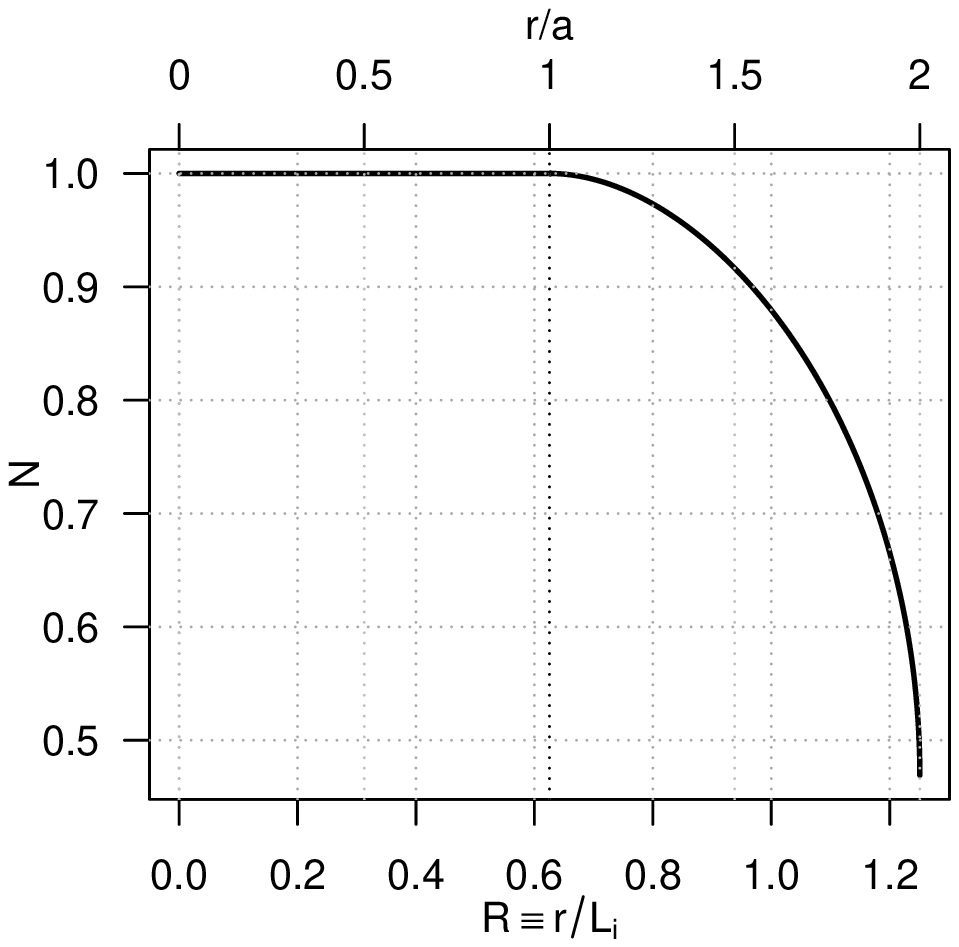}
\caption{\label{fig:qn} Numerical quasi-neutral solutions for $U$ (left panel, solid curve), $\Phi$ (left panel, dashed curve), and $N$ (right panel) with $A = 0.6254204$.}
\end{figure*}

We simultaneously compute $N$ from eq.\ (\ref{eq:NqnII}), and $\Phi$ as $\ln \, (1/N)$.
Given the boundary condition that $U = 1$ at $R = 2A$, the choice of $A$ determines $N$ and hence $\Phi$ at the interface. Since $\Phi = 0$ there, we must then choose $A$ to make $N = 1$ at $R = A$. By numerical trial and improvement we find this critical $A$ value is 0.62542, so the total radius of the quasi-neutral system is $1.251 L_i$, and $N$ at its outermost edge is 0.468. Figure \ref{fig:qn} presents this canonical solution for $U$ and $N$.

The quasi-neutral solution does not hold near the system's outer wall, where the plasma approximation fails --- in reality a sheath must form there, so $n_i$ ceases to be approximately $n_e$. However, in the $\Lambda \rightarrow 0$ limit the sheath is vanishingly thin and the quasi-neutral solution gives excellent results almost everywhere. The quasi-neutral solution therefore provides a rough check on numerical solutions with small $\Lambda$.

\section{Numerical solution for region II with finite Debye length}

Because $\Lambda$ is nonzero in practice, realistic solutions of eqs.\ (\ref{eq:normcont})--(\ref{eq:normPois}) differ from the quasi-neutral solutions, and the equations must be re-solved numerically. To accomplish this we apply the shooting method. In each iteration of the shooting method we solve the three differential equations by integrating outwards from near the interface with the midpoint method.

\begin{figure*}
\includegraphics[width=0.49\textwidth]{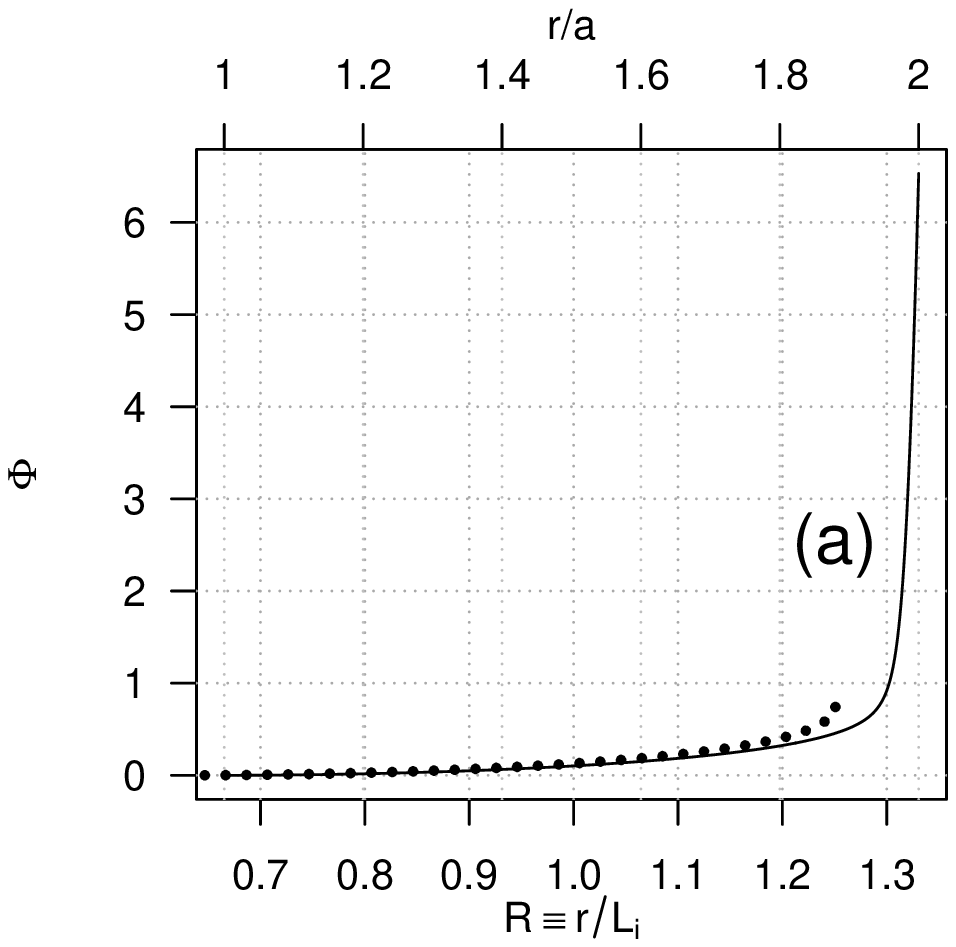}
\includegraphics[width=0.49\textwidth]{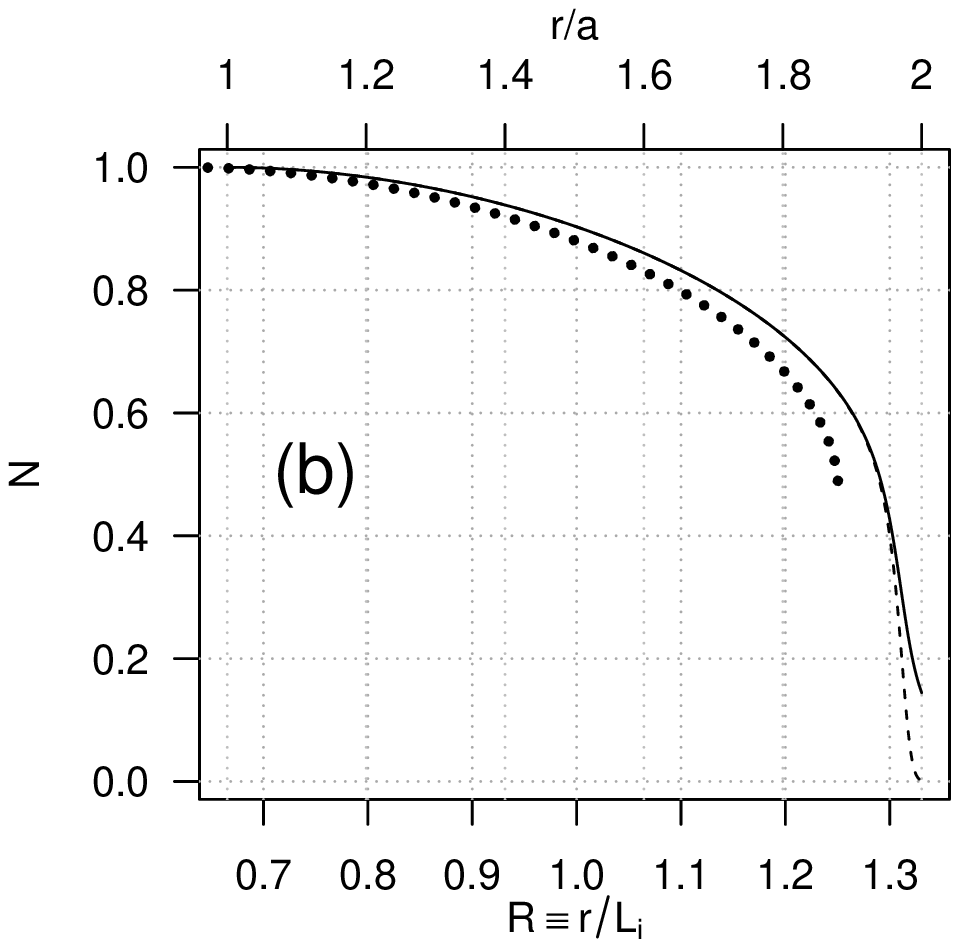}
\includegraphics[width=0.49\textwidth]{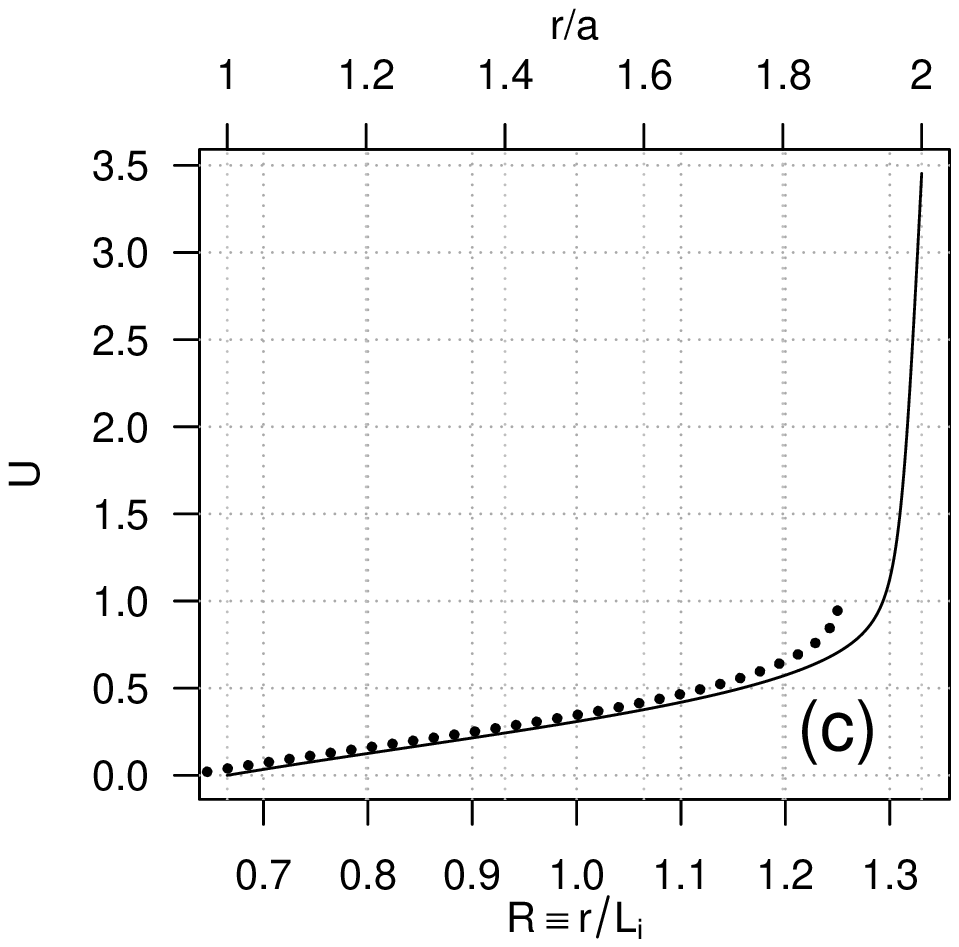}
\includegraphics[width=0.49\textwidth]{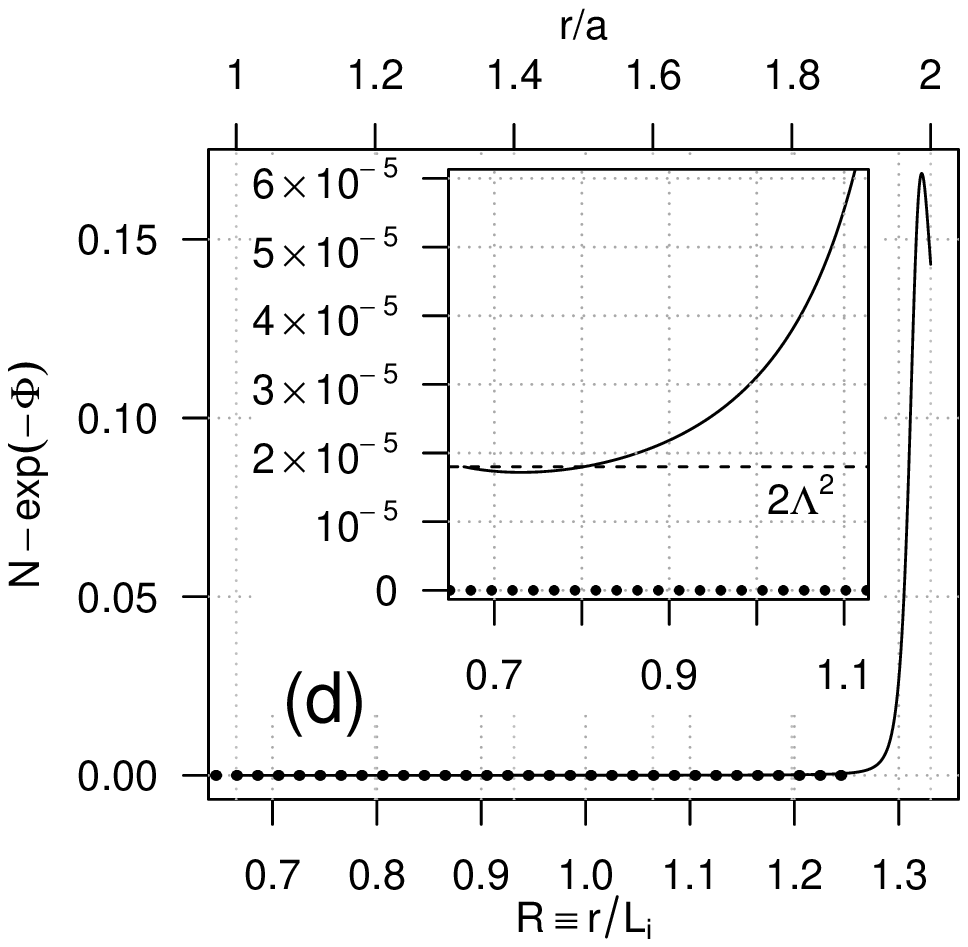}
\caption{\label{fig:fin-lam} Numerical region II solution for $\Lambda = 0.003$ (thin curves) compared to the canonical quasi-neutral solution (thick dotted curves). We augment the default choice of axis scale ($R$, on the lower axes) with the alternative choice (on the upper axes) of $r$ normalized by $a$'s value when $\Lambda = 0.003$. In plot (b), the thin solid curve represents the ion density while the thin dashed curve represents the electron density $\exp(-\Phi)$.}
\end{figure*}

The normalized system has the four dimensionless parameters $m_i / m_e$, $G_{II}$, $\Lambda$, and $A$; once any three are set the fourth is then constrained to a particular value by the outer wall boundary condition.
We fix $m_i / m_e = 736744$ (corresponding to Hg$_2^+$ ions, the ion species considered in our previous papers), $G_{II} = 1$, and $\Lambda = 0.003$, then find $A$ by trial and improvement, converging to an $A$ for which eq.\ (\ref{eq:equfluxbc}) is approximately satisfied. Figure \ref{fig:fin-lam} presents this solution of the system. We find that the region II plasma has a sheath adjacent to the outer wall, but has minimal net space charge closer to the plasma's central axis. Near the interface ($R \approx A$) the numerical solution is similar to the quasi-neutral solution, but the two solutions gradually deviate farther out.

A subtle but important feature is a discontinuity in $N$ at the interface. In region I, $N$ is exactly unity. However, for our current parameters, $N$ discontinuously jumps from 1 to 1.000018 at the beginning of region II.
The electron density is continuous at the interface because it is given by the Boltzmann relation, and we cannot have discontinuities of potential. This argument does not apply to the cold positive ions. The ion density must change in order to produce a very small space charge density (a divergence of the electric field) in the outer plasma. We can note that this does not imply a layer of charge on the boundary, the electric field is continuous.
Consequently, all of region II is a positive space charge region, not just its sheath, but that space charge is too subtle to be visible near the interface unless one looks closely (figure \ref{fig:fin-lam}, inset).

\begin{figure}
\includegraphics[width=0.49\textwidth]{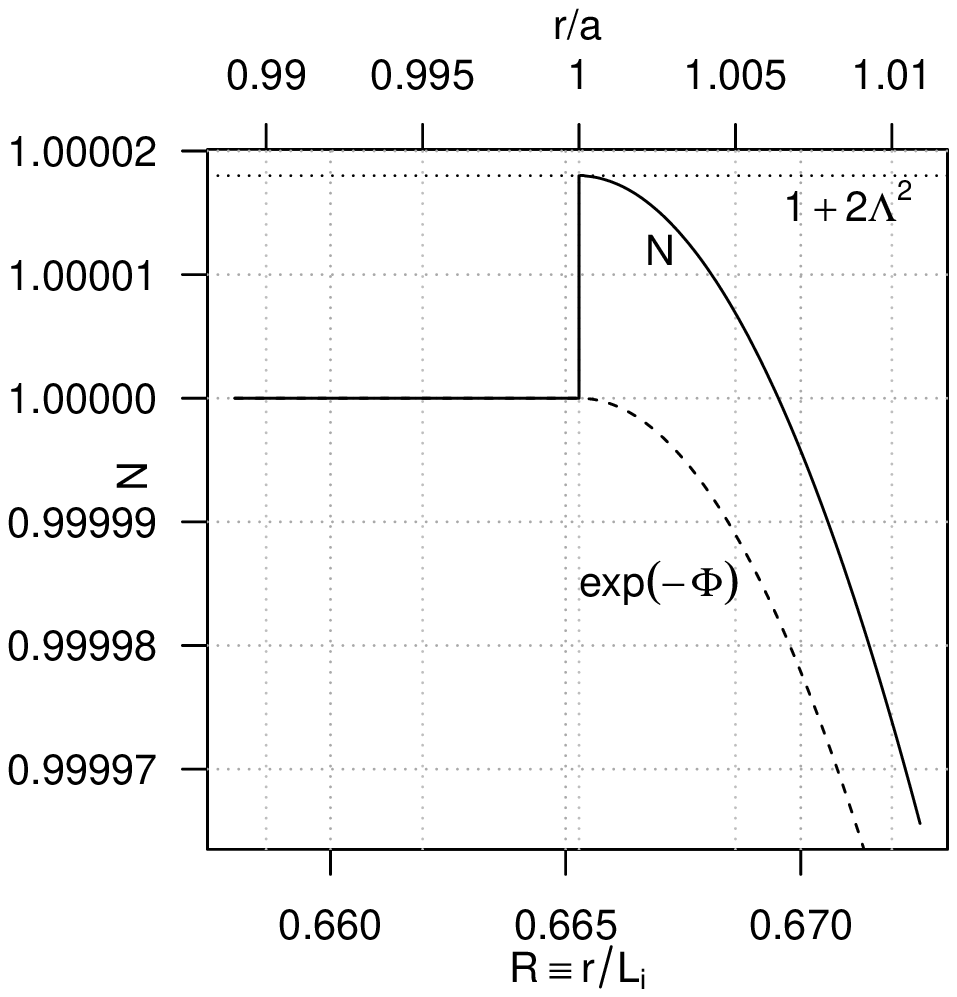}
\caption{\label{fig:fin-lam-discon} Detail of the ion density discontinuity at the interface between regions I and II in the $\Lambda = 0.003$ case with Hg$_2^+$ ions and $G = 1$.}
\end{figure}

$N$'s discontinuity is not an artifact. Even under quasi-neutrality, $N$ and $\Phi$ have nonzero curvature at the interface, which in physical terms implies a nonzero space charge density. To estimate that density, we rearrange eq.\ (\ref{eq:NUII}) for $U$ and substitute into the quasi-neutral ion momentum equation. After some algebra that gives
\begin{multline}
-N \frac{\dee N}{\dee R} = \frac{R^2 - A^2}{2R}
+ \frac{R^2 - A^2}{4 R^3 N} \Bigg( \left( R^2 + A^2 \right) N \\
- \left( R^2 - A^2 \right) R \frac{\dee N}{\dee R} \Bigg)
\label{eq:qnNODE}
\end{multline}
At $R = A$, $\dee N / \dee R = 0$ and $R^2 - A^2 = 0$, so at that $R$ all terms in eq.\ (\ref{eq:qnNODE}) are nil. But the terms do not all have the same dependence on $R-A$. Unlike the other terms, the very last term is a product of $\dee N / \dee R$ \emph{and} of $\left( R^2 - A^2 \right)$ \emph{squared}, so for $0 < (R - A) \ll 1$ it is negligible relative to the other terms. We therefore drop this term to get
\begin{equation}
-N \frac{\dee N}{\dee R} \approx \frac{R^2 - A^2}{2R}
+ \frac{R^4 - A^4}{4 R^3}
\end{equation}
which has the specific solution
\begin{equation}
N \approx \frac{1}{2}
\sqrt{4 \left( N_0^2 + A^2 \right) - 3 R^2
- \frac{A^4}{R^2} + 4 A^2 \ln \frac{R}{A}}
\label{eq:interfN}
\end{equation}
where $N_0$ is $N$'s value at the interface ($R = A$). Under quasi-neutrality, $\Phi = - \ln \, N$, so we substitute the $N$ solution into that expression for $\Phi$ and take $\Phi$'s Taylor series about $R = A$ to second order:
\begin{align}
\Phi & \approx -\ln \, N_0 + \frac{(R - A)^2}{N_0^2} \label{eq:interfPhi} \\
\Rightarrow \frac{\dee^2 \Phi}{\dee R^2} & \approx \frac{2}{N_0^2}
\end{align}
$\Phi$'s curvature is not going to be much different for sufficiently small $\Lambda$, so we now let $\Lambda$ be finite and reintroduce Poisson's equation:
\begin{equation}
\frac{\dee^2 \Phi}{\dee R^2}
= \frac{N - \exp(-\Phi)}{\Lambda^2} - \frac{1}{R} \frac{\dee \Phi}{\dee R}
\approx \frac{2}{N_0^2}
\end{equation}
At the interface, $N = N_0$ and $\Phi = \dee \Phi / \dee R = 0$, so
\begin{equation}
\frac{N_0 - 1}{\Lambda^2} \approx \frac{2}{N_0^2}
\end{equation}
a cubic equation with the solution
\begin{equation}
N_0 \approx \frac{1 + \nu + 1/\nu}{3}
\end{equation}
where
\begin{equation}
\nu \equiv \sqrt[3]{ 1 + 3 \Lambda \sqrt{6 + 81 \Lambda^2} + 27 \Lambda^2 }
\end{equation}
For all $\Lambda > 0$, $\nu$ and hence $N_0$ are greater than 1.
It follows that $N$ jumps discontinuously from 1 to $N_0$ where region I becomes region II, and the size of the jump is the net space charge density at the interface.
(Our numerical solver uses this fact to compute boundary conditions for the region II equations.)
When $\lambda_D \ll L_i$, $N_0 \approx 1 + 2 \Lambda^2$, so the net space charge density at the interface is $\approx 2 \Lambda^2$ (figure \ref{fig:fin-lam-discon}).

It is straightforward to consider the solution's spatial dependence near the interface rather than its $\Lambda$ dependence. Taking Taylor series of eq.\ (\ref{eq:interfN}) about $R=A$,
\begin{equation}
N = N_0 - \frac{(R-A)^2}{N_0} + \mathcal{O} \left( (R-A)^3 \right)
\end{equation}
In the limit of $\Lambda$ vanishing outright, $N_0 \rightarrow 1$. We then have $N \approx 1 - (R-A)^2$, $\Phi \approx (R-A)^2$ from eq.\ (\ref{eq:interfPhi}), and, combining the continuity equation with the fact that $U = 0$ and $G$ jumps to 1 at the interface, $\dee U / \dee R \approx 1/N \approx 1$, so $U \approx (R-A)$.
This matches the behaviour of the curves in figure \ref{fig:qn} just beyond the interface: $N$ falls parabolically, $\Phi$ grows parabolically, and $U$ rises nearly linearly.

\begin{figure}
\includegraphics[width=0.49\textwidth]{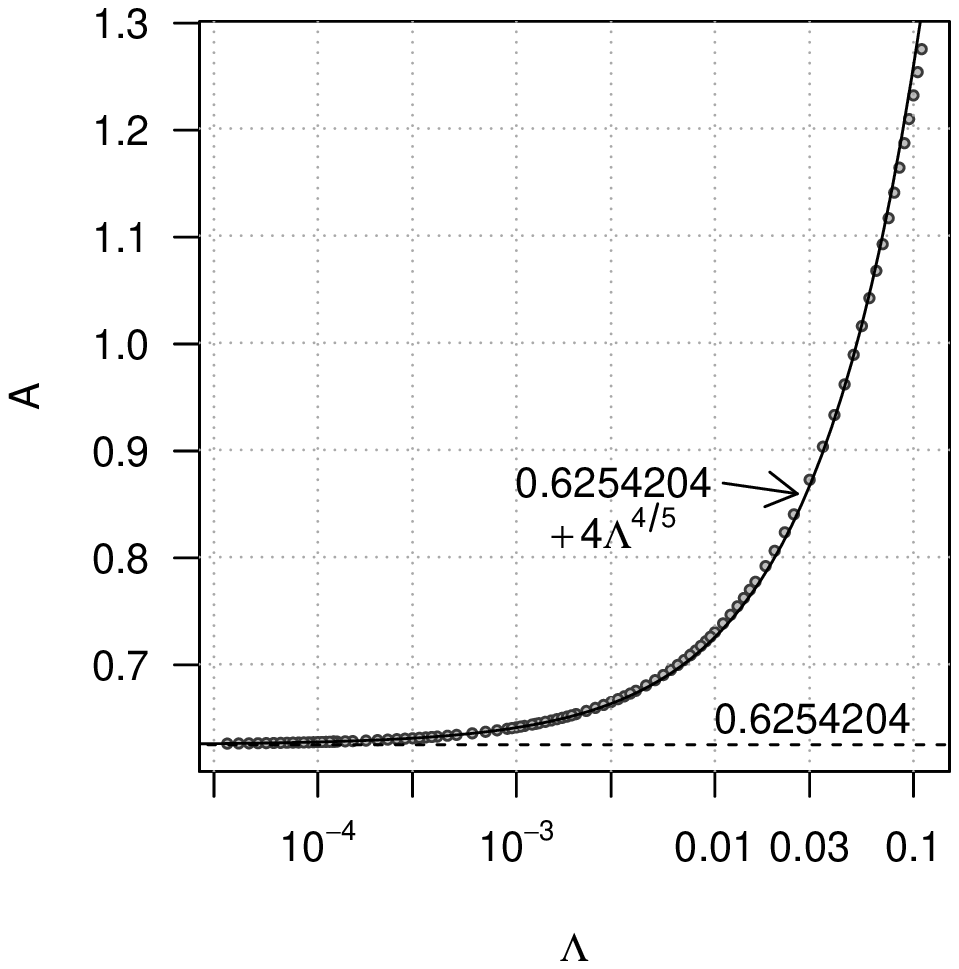}
\caption{\label{fig:A-of-lam} $A$ as a function of $\Lambda$ given $G = 1$ and Hg$_2^+$ ions.}
\end{figure}

Just as the quasi-neutral case casts light on systems with $0 < \Lambda \ll 1$, our numerical solutions of the latter support our quasi-neutral result. Computing $A$ for assorted small values of $\Lambda$ (with $G$ and $m_i / m_e$ fixed at 1 and 736744 respectively), it's evident that as $\Lambda$ shrinks, the resulting $A$ values converge on 0.62542 (figure \ref{fig:A-of-lam}), the canonical $A$ value found earlier by solving the quasi-neutral equations.
For $\Lambda \ll 1$, $A$ is greater than the canonical quasi-neutral $A$ by about $4 \Lambda^{4/5}$.

\begin{figure}
\includegraphics[width=0.49\textwidth]{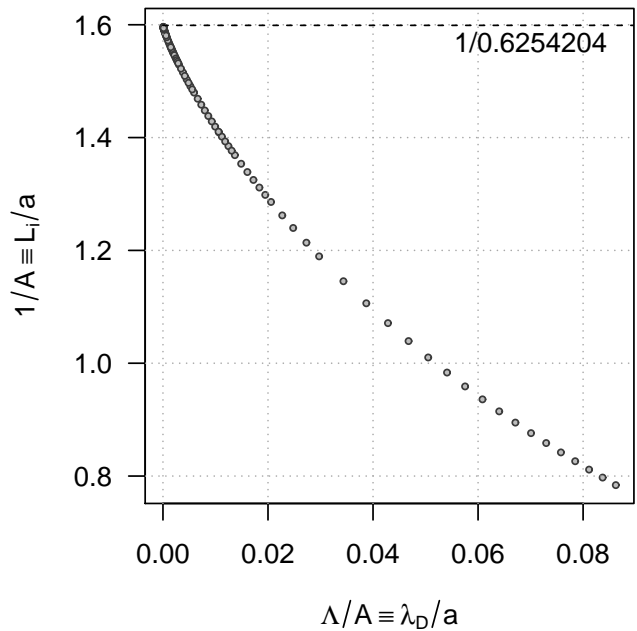}
\caption{\label{fig:one-over-A} $1/A$ as a function of $\Lambda / A$ given $G = 1$ and Hg$_2^+$ ions. The dashed line is $1/A$ in the $\Lambda \rightarrow 0$ limit.}
\end{figure}

Figure \ref{fig:one-over-A} displays the same information in a different manner. The quantity $1/A$ illustrates that the ionization length is comparable to the plasma dimension, but depends on $\lambda_D$'s magnitude in comparison with the latter.

\section{Conclusion}

An interesting result of the present investigation is that a uniform field-free plasma forms in the inner region. This is an extremely rare occurrence in plasma physics, the other example known to us being that of a thermally produced caesium plasma in a cavity \cite{Phelps76}.
It is clear that further calculations could readily be carried out. An example would be to use a Tonks-Langmuir model \cite{Tonks29} in which the positive ions have a distribution of velocities in the radial direction. Collisions with neutral atoms could also be included in the analysis. A process not included in the model is that electrons produced by irradiation may then contribute to further ionization in collisions with neutral atoms.

We have also confined ourselves in this paper to the system's steady-state solutions, but one could carry out a transient analysis which accounted for the initial flow of electrons into region I as the plasma establishes equilibrium. This electron flow would produce an electric field drawing the ions in after the electrons, a phenomenon akin to the expansion of a plasma into a vacuum \cite{Crow75}. Modelling transient effects could also quantify how the plasma's behaviour adjusts the fundamental, observable plasma parameters ($n_0$, $T_e$, and $g_{II}$) to move $\Lambda$ to its steady-state value, even if the experimenter's initial conditions give $\Lambda$ a different starting value.

Calculations could be carried out for a less abrupt change between regions I and II as briefly discussed in our earlier paper \cite{Franklin13}, and for different ratios of the two regions' radii.
We made our choice of positive ion, i.e.\ Hg$_2^+$, for historical reasons given in the same paper \cite{Franklin13}. The ratio of this ion mass to the electron mass enters the theory through the boundary condition at the wall. In effect it determines only the width of the sheath and the voltage across it.
It is of interest to note that the ion velocity on entering the sheath is about equal to the Bohm velocity, whereas in our earlier model, which had ionization in the inner region, considerably higher velocities were found \cite{Franklin13}.
An even more striking difference between that model and the current model is the behaviour of their solutions near the interface. Our earlier model produced a double layer centred on the interface, while our current model has a jump discontinuity in density.

Our previous paper gave an indication of the equipment needed for experimental work in this field \cite{Franklin13}. The model system we present here should be relatively easy to realize in the laboratory, especially compared to our original model system, and could be a way to make uniform, stationary plasmas on command. To accomplish this, annular UV illumination of a cylindrical container of mercury gas should suffice. We are hopeful that both experimental work and further theoretical work shall be carried out to make further progress in this vein.

\section{Acknowledgement}

We are indebted to Professor Raoul Franklin for his lead in initiating this particular programme of work.


\begin{thebibliography}{6}

\bibitem{Franklin13}
R.\ N.\ Franklin, J.\ E.\ Allen, D.\ M.\ Thomas, and M.\ S.\ Benilov, \textit{Phys.\ Plasmas} \textbf{20}, 123508 (2013).

\bibitem{Benilov14}
M.\ S.\ Benilov and D.\ M.\ Thomas, \textit{Phys.\ Plasmas} \textbf{21}, 043501 (2014).

\bibitem{Phelps76}
A.\ D.\ R.\ Phelps and J.\ E.\ Allen, \textit{Proc.\ R.\ Soc.\ Lond.\ A} \textbf{348}, 221 (1976).

\bibitem{Bohm49}
D.\ Bohm, \textit{The characteristics of electrical discharges in magnetic fields}, chapter 3 (McGraw-Hill, New York, 1949).

\bibitem{Tonks29}
L.\ Tonks and I.\ Langmuir, \textit{Phys.\ Rev.} \textbf{34}, 876 (1929).

\bibitem{Crow75}
J.\ E.\ Crow, P.\ L.\ Auer, and J.\ E.\ Allen, \textit{J.\ Plasma Phys.} \textbf{14}, 65 (1975).

\end{thebibliography}
\end{document}